\documentclass[a4paper]{article}
\usepackage[psamsfonts]{amssymb}
\usepackage{amsmath}
\usepackage{cite}
\usepackage{epsfig}
\usepackage{graphicx}
\date{}

\author{M. Alimohammadi\footnote{alimohmd@ut.ac.ir}\ \ and H.
Behnamian
\\ {\small Department of Physics, University of Tehran,}
\\ {\small North Karegar Avenue, Tehran, Iran.}}
\title{Remarks on generalized scalar-tensor models of dark energy }
\begin{document}
\maketitle
\begin{abstract}
The generalized scalar-tensor models with Lagrangian
$F(\phi,R)-U(\phi)(\nabla\phi)^2$ are considered. It is shown that
the phantom-divide-line crossing and the deceleration to
acceleration transition generally occurr in these models. Two
specific examples, the coupled quintessence model and the
Brans-Dicke model are considered. For the first example, it is
shown that for the models with $\xi>3/16$, the $\omega=-1$
transition exists. This is verified numerically for some special
cases. For the Brans-Dicke model, it is shown that the transition
does not occur, a result which can be verified by using the exact
solution of this model. Finally the contribution of quantum
effects on these phenomena is investigated. It is shown that for
some special cases where the $\omega=-1$ transition is classically
forbidden, the quantum effects can induce transition. The
$\xi=1/6$ of coupled quintessence model is an example of this. The
quantum effects are described via the account of conformal
anomaly.
\end{abstract}
\section{Introduction}
Nowadays based on various observational data, it has been verified
that our universe is now in an accelerating phase~\cite{Riess}.
The source of this accelerated expansion, which is known as dark
energy, is believed to compose nearly 70$\%$ of the present
universe. The precise nature of dark energy is not yet known but
all of the models describing the dark energy have a common
property: they produce the repulsive force, or in other words,
they have the negative pressure. Introducing the equation of state
parameter by $\omega=p/\rho$, where $p$ is the pressure and $\rho$
is the energy density, it must satisfy $\omega<-1/3$, in
Friedmann-Robertson-Walker (FRW) background metric, to ensure the
accelerated expansion.

One of the interesting features of dark energy is the dynamical
behavior of equation of state parameter $\omega$. This is because
some astrophysical data slightly favor an evolving dark energy and
show a recent $\omega=-1$, the so-called phantom-divide-line,
crossing~\cite{Huterer}. To describe the behavior of $\omega$,
many dynamical dark energy models have been introduced. The
simplest of these models are the single component scalar models,
including the quintessence model, which consists of a normal
scalar field~\cite{liddel}, and the phantom model, which is a
scalar field theory with unusual negative kinetic
energy~\cite{caldwell}. None of these models can describe the
$\omega=-1$ crossing. In quintessence model $\omega$ is always
$\omega>-1$, and in phantom model $\omega$ always satisfies
$\omega<-1$. A possible way to overcome this problem is to
consider the quintom model~\cite{feng}, the model consists of one
quintessence and one phantom fields. It can be shown that the
transition from $\omega>-1$ to $\omega<-1$ always occurs in the
quintom model with slowly-varying potential~\cite{mohseni}. The
interaction of single-component scalar field with background
(dark) matter can also induce this transition~\cite{chimento}.

An alternative candidate of dark energy are the models known as
modified gravities, in which it is postulated that the gravity is
being nowadays modified by some extra terms which grow when the
curvature decreases. The $f(R)$ gravity, whose action is a general
function $f(R)$ of the Ricci scalar $R$, is the most famous
modified gravity theory~\cite{nojiri1}. The $f(G)$ gravity and its
generalization $F(R,G)$ gravity, where $G$ being the Gauss-Bonnet
invariant, are also introduced~\cite{nojiri2}. Several aspects of
these models, including their behavior under $\omega=-1$
transition, have been studied recently. It has been shown that the
phantom-divide-line crossing and the deceleration to acceleration
transition generally occur in $F(R,G)$ models. The contribution of
quantum effects on these phenomenon has been also
obtained~\cite{alimohammadi1}.

An important class of modified gravity theories is the
scalar-tensor theories. In these models , there exists a scalar
field $\phi$, besides the usual space-time metric $g_{\mu\nu}$, to
describe the gravitational interaction. The earliest type of these
models is the Jordan-Brans-Dicke theory with the following action,
in the so-called Jordan frame,~\cite{brans}
\begin{equation}\label{1}
S=\frac{1}{2}\int {\rm d}^{4}x\sqrt{-g}{}\hspace{1ex}\left[ \phi
R-\frac{{\rm{w}}}{\phi}(\nabla\phi)^2-V(\phi) \right]+S_m.
\end{equation}
Here we use the units in which $\hbar=c=8\pi G=1$. 'w' is a
dimensionless parameter and $S_m$ is the action of dustlike
matter. The scalar field $\phi$ plays the role of the inverse of
the gravitational coupling, which is constant in general
relativity but in Jordan-Brans-Dicke theory is variable. For the
case where $V(\phi)=0$, the model is called the Brans-Dicke
theory.

The generalization of the Jordan-Brans-Dicke theory is the
scalar-tensor theory which is described, in the Jordan frame, by
the action~\cite{bergmann}
\begin{equation}\label{2}
S=\frac{1}{2}\int {\rm d}^{4}x\sqrt{-g}{}\hspace{1ex}\left[
F(\phi) R-U(\phi)(\nabla\phi)^2-V(\phi) \right]+S_m.
\end{equation}
Several aspects of scalar-tensor models have been studied,
including their behavior as a dynamical system~\cite{billyard},
the gravitational wave and inflation in this
framework~\cite{will}, and their perturbative
aspects~\cite{boisseau,tsujikawa}. Two review articles on this
subject are~\cite{faraoni}.

The most general action in the Jordan frame, based on the
scalar-tensor theory, is the following:
\begin{equation}\label{3}
S=\frac{1}{2}\int {\rm d}^{4}x\sqrt{-g}{}\hspace{1ex}\left[
F(\phi,R)-U(\phi)(\nabla\phi)^2 \right]+S_m,
\end{equation}
which is called the generalized scalar-tensor (ST)
theories~\cite{tsujikawa,faraoni}. The scalar theories, i.e.
quintessence and phantom models, $f(R)$ gravity models,
Jordan-Brans-Dicke theory (\ref{1}), and the scalar-tensor models
(\ref{2}) are special examples of action (\ref{3}). It is must be
noted that the terminology is not fixed in scalar-tensor theories.
The term "generalized scalar-tensor models" has been used for less
general actions than eq.(\ref{3}), even for action (\ref{2}) with
$F(\phi)=\phi$.

The present paper is devoted to the study of generalized ST
theories. We want to study the conditions under which the
$\omega=-1$ crossing occurs in generalized scalar-tensor theories.
The condition of transition between deceleration and acceleration
phases is also studied. We show that, through some specific
examples, the phantom-divide-line crossing can not occur for some
models. It is shown that by considering the quantum contributions,
the situation becomes different. In fact, some of the forbidden
transitions become allowed, i.e. the $\omega=-1$ transitions, for
these cases, are quantum induced. The same phenomenon has been
seen for quintessence and phantom theories~\cite{alimohammadi2}.
The quantum effects are described via the account of conformal
anomaly, reminding about anomaly-driven
inflation~\cite{starobinsky}. The contribution of the conformal
anomaly in energy conditions and Big Rip of phantom models has
been discussed in~\cite{nojiri3}.

The scheme of the paper is as follows. In section 2, the Friedmann
equations of generalized ST theories for FRW space-time are
obtained. By perturbative solving of these equations around the
$\omega=-1$ and $\ddot{a}=0$ transition points in sections 3 and
4, respectively, the conditions of occurrence of these crossings
are obtained. In section 5, we consider the explicit example of
coupled quintessence model in which there exists a free parameter
$\xi$. It is shown that the $\omega=-1$ transition occurs if
$\xi>3/16$. It is seen that the numerical solving of Friedmann
equations verifies our perturbative results. In section 6, the
Brans-Dicke model is considered as a second example. Our method
shows no transition for this model, a result which can be verified
by exact solution of this model. Finally in section 7, we
introduce the quantum contributions (comes from conformal anomaly)
to this problem, and show that it can induce the $\omega=-1$
transition for special cases which are classically forbidden. For
example for $\xi=1/6$ case of the coupled quintessence model, in
which the transition is classically forbidden, the quantum
phenomena induce the $\omega=-1$ transition.

\section{The generalized scalar-tensor models}

Consider the generalized scalar-tensor gravity with action
(\ref{3}). Varying this action with respect to the metric
$g_{\mu\nu}$ results in~\cite{faraoni}
\begin{equation}\begin{split}\label{4}
&( R_{\mu\nu}-\nabla_\mu\nabla_\nu+g_{\mu\nu}\Box) F_R(\phi,R)-\frac{1}{2}g_{\mu\nu}\left[ F (\phi,R)\right.\\
&\left.-U(\phi)(\nabla\phi)^2\right]-U(\phi)\nabla_\mu\phi\nabla_\nu\phi=T_{\mu\nu}.
\end{split}\end{equation}
In this equation, $T_{\mu\nu}$ is the energy-momentum tensor of
matter field
\begin{equation}\label{5}
T_{\mu\nu}=\frac{2}{\sqrt{-g}}\frac{\delta S_m}{\delta
g_{\mu\nu}},
\end{equation}
and
\begin{equation}\label{6}
F_R(\phi,R)=\frac{\partial F(\phi,R)}{\partial R}\hspace{4ex} ,
\hspace{4ex}F_\phi(\phi,R)=\frac{\partial
F(\phi,R)}{\partial\phi}.
\end{equation}
Varying the action with respect to the scalar field $\phi$,
results in
\begin{equation}\label{7}
F_\phi(\phi,R)+2U(\phi)\Box\phi+2\nabla^{\mu}\phi\nabla_{\mu}U(\phi)
-\frac{dU(\phi)}{d\phi}(\nabla\phi)^{2}=0.
\end{equation}
For the case where $U(\phi)=\eta=$constant, eq.(\ref{7}) reduces
to the corresponding equation of motion in~\cite{faraoni}.

A spatially flat FRW space-time in comoving coordinates
$(t,x,y,z)$ is defined through
\begin{equation}\label{8}
{\rm d}s^{2}=-{\rm d}t^{2}+a^2(t)({\rm d}x^{2}+{\rm d}y^{2}+{\rm
d}z^{2}) ,
\end{equation}
where $a(t)$ is the scale factor. For this metric, the $(t,t)$
component of the evolution equation (\ref{4}) for homogenous
scalar field $\phi(t)$ becomes:
\begin{equation}\begin{split}\label{9}
&3H^{2}F_R(\phi,R)=\rho_{m}+\frac{1}{2}U(\phi)\dot{\phi}^{2}-3HF_{RR}(\phi,R)\dot{R}\\
&-3HF_{R\phi}(\phi,R)\dot{\phi}-\frac{1}{2}F(\phi,R)+\frac{1}{2}RF_{R}(\phi,R).
\end{split}\end{equation}
$H(t)={\dot{a}(t)}/{a(t)}$ is the Hubble parameter and $\rho_{m}$
is the matter energy density with the evolution equation
\begin{equation}\label{10}
\dot{\rho}_m+3H(\rho_m+p_{m})=0.
\end{equation}
$R$ is expressed in terms of $H$ as following
\begin{equation}\label{11}
R=6(\dot{H}+2H^{2}).
\end{equation}
The $(x,x)$ component of eq.(\ref{4}), when eq.(\ref{9}) is
subtracted from it, becomes
\begin{equation}\begin{split}\label{12}
-2\dot{H}F_{R}(&\phi,R)+HF_{RR}(\phi,R)\dot{R}+HF_{R\phi}(\phi,R)\dot{\phi}
-F_{RRR}(\phi,R)\dot{R}^{2}
-2F_{RR\phi}(\phi,R)\dot{R}\dot{\phi}\\
\\ &-F_{RR}(\phi,R)\ddot{R}-F_{R\phi\phi}(\phi,R)
\dot{\phi}^{2}-F_{R\phi}(\phi,R)\ddot{\phi}
-U(\phi)\dot{\phi}^{2}=\gamma_{m}\rho_{m}.
\end{split}\end{equation}
In above equation, $\gamma_{m}$ is defined by
$\gamma_{m}=1+\omega_{m}$, where $\omega_{m}={p_{m}}/{\rho_{m}}$,
and the subscripts denote differentiations, e.g.
\begin{equation}\label{13}
F_{RR\phi}(\phi,R)=\frac{\partial^{3}F(\phi,R)}{\partial^{2}R\,\partial\phi}.
\end{equation}
Finally the scalar field equation of motion (\ref{7}) in FRW
metric becomes:
\begin{equation}\label{14}
F_{\phi}(\phi,R)-2U(\phi)(\ddot{\phi}+3H\dot{\phi})-\dot{U}(\phi)\dot{\phi}=0.
\end{equation}
In above equation $\dot{U}=({\rm d}U/{\rm d}\phi)\dot{\phi}$.
Equations (\ref{9}), (\ref{10}), (\ref{12}) and (\ref{14}) are the
Friedmann equations of generalized scalar-tensor models. Note that
these four equations are not independent, in fact, taking the time
derivative of eq.(\ref{9}), adding it eq.(\ref{12}), multiplied by
$3H$,  and using eq.(\ref{14}), result in eq.(\ref{10}). So there
are three independent Friedmann equations. In the case of ordinary
scalar-tensor models where $F(\phi,R)=RF(\phi)-V(\phi)$, the
resulting Friedmann equations lead to the corresponding ones in
scalar-tensor theories.

\section{The phantom-divide-line crossing}

For the ordinary dark energy models and in the context of Einstein
gravity, the equation of state parameter $\omega=p/\rho$ is
obtained via $\omega=-1-2\dot{H}/3H^{2}$. For other theories,
including the usual and generalized ST theories, the effective
equation of state parameter $\omega_{\rm{eff}}$ is also defined
through~\cite{nojiri2,tsujikawa}
\begin{equation}\label{15}
\omega_{\rm{eff}}=-1-\frac{2}{3}\frac{\dot{H}}{H^{2}}.
\end{equation}
So $\omega_{\rm{eff}}$ crosses the phantom-divide-line, i.e.
$\omega_{\rm{eff}}=-1$, if $H(t)$ has a relative extremum at some
time $t=t_{0}$. Restricting ourselves to $t-t_{0}\ll h_{0}^{-1}$,
where $h_{0}=H(t_{0})$ and $h_{0}^{-1}$ is of order of the age of
our universe, the Hubble parameter can be expanded as
\begin{equation}\label{16}
H(t)=h_{0}+h_{1}(t-t_{0})^{\alpha}+h_{2}(t-t_{0})^{\alpha+1}+O((t-t_{0})^{\alpha+2}),
\end{equation}
in which $\alpha\geq 2$ is the order of the first nonvanishing
derivative of $H(t)$ at $t=t_{0}$, and $h_{1}=(1/\alpha
!)H^{(\alpha)}(t_{0})$. $H^{(n)}(t_{0})$ is $n$th derivative of
$H(t)$ at $t=t_{0}$ . The transition occurs from $\omega>-1$ to
$\omega<-1$ regions when $\alpha$ is an even positive integer and
$h_{1}>0$. For $h_{1}<0$, and even integer $\alpha$, the system
goes from $\omega<-1$ to $\omega>-1$ at $t=t_{0}$. We now examine
whether there is any solution for Friedmann eqs.(9), (10), and
(14), if $H(t)$ is given by (16). If for an even positive integer
$\alpha$, we can find a nonzero value for $h_{1}$, then the
possibility of $\omega=-1$ transition of generalized ST theories
is proved.

We first consider the Friedmann equation (\ref{9}):
\begin{equation}\label{17}
H^{2}F_{R}(\phi,R)\equiv\beta(t),
\end{equation}
in which
\begin{equation}\begin{split}\label{18}
&\beta(t)=\frac{1}{3}[\hspace{.1cm}\rho_{m}+\frac{1}{2}U(\phi)\dot{\phi}^{2}-3HF_{RR}(\phi,R)\dot{R}\\
-&3HF_{R\phi}(\phi,R)\dot{\phi}-\frac{1}{2}F(\phi,R)+\frac{1}{2}RF_{R}(\phi,R)\hspace{.1cm}].
\end{split}\end{equation}
Expanding both sides of eq.(\ref{17}) around $t_{0}\equiv0$, one
finds, up to order $t^{2}$,:
\begin{equation}\label{19}
h_{0}^{2}F_{R}(0)=\beta(0),
\end{equation}
\begin{equation}\label{20}
h_{0}^{2}(F_{RR}\dot{R}+F_{R\phi}\dot{\phi})_{t=0}=\dot{\beta}(0),
\end{equation}
and
\begin{equation}\begin{split}\label{21}
\frac{1}{2}h_{0}^{2}(F_{RRR}\dot{R}^{2}&+2F_{RR\phi}\dot{R}\dot{\phi}
+F_{RR}\ddot{R}+F_{R\phi\phi}\dot{\phi}^{2}+F_{R\phi}\ddot{\phi})_{t=0}\\
&+2h_{0}h_{1}F_{R}(0)\delta_{\alpha,2}=\frac{1}{2}\ddot{\beta}(0).
\end{split}\end{equation}
The second Friedmann equation (\ref{10}), at zero order, leads to
\begin{equation}\label{22}
\dot{\rho}_{m}(0)+3h_{0}\gamma_{m}\rho_{m}(0)=0,
\end{equation}
and the last equation of motion (\ref{14}), up to order $t$,
results in:
\begin{equation}\label{23}
F_{\phi}(0)-2U(0)(\ddot{\phi}+3h_{0}\dot{\phi})_{t=0}-U_{\phi}(0)\dot{\phi}^{2}(0)=0,
\end{equation}
and
\begin{equation}\begin{split}\label{24}
F_{\phi\phi}(0)\dot{\phi}&(0)-U_{\phi}(0)(4\dot{\phi}\ddot{\phi}
+6h_{0}\dot{\phi}^{2})_{t=0}-2U(0)(\dddot{\phi}+3h_{0}\ddot{\phi})_{t=0}\\
&-U_{\phi\phi}(0)\dot{\phi}^{3}(0)+12h_{1}F_{R\phi}(0)\delta_{\alpha,2}=0.
\end{split}\end{equation}
Using eqs.(\ref{11}), (\ref{22}) and (\ref{23}), the
eqs.(\ref{19}) and (\ref{20}) lead to:
\begin{equation}\begin{split}\label{25}
&3h_{0}^{2}F_{R}(0)+\rho_{m}(0)+\frac{1}{2}U(0)\dot{\phi}^{2}(0)-\frac{1}{2}F(0)\\
&-3h_{0}F_{R\phi}(0)\dot{\phi}(0)-36h_{0}h_{1}F_{RR}(0)\delta_{\alpha,2}=0,
\end{split}\end{equation}
and
\begin{equation}\begin{split}\label{26}
6[ 3h_{0}(h_{2}&+h_{0}h_{1})F_{RR}+12h_{0}h_{1}^{2}F_{RRR}
+2h_{0}h_{1}F_{RR\phi}\dot{\phi}]_{t=0}\delta_{\alpha,2}
+18h_{0}h_{1}F_{RR}(0)\delta_{\alpha,3}\\&
+\frac{1}{2}[{}\hspace{.1cm}\gamma_{m}\rho_{m}-h_{0}F_{R\phi}\dot{\phi}
+U(\phi)\dot{\phi}^{2}+F_{R\phi\phi}\dot{\phi}^2
+F_{R\phi}\ddot{\phi}{}\hspace{.1cm}]_{t=o}h_{0}=0,
\end{split}\end{equation}
respectively. Note that at $t=0$, $R=R(0)=12h_{0}^{2}$.

Let us first consider the case $\alpha=2$. The eqs.(\ref{22}),
(\ref{25}) and (\ref{26}) then result in $h_{0}$, $h_{1}$ and
$h_{2}$, respectively. $h_{0}$ and $h_{1}$ are
\begin{equation}\label{27}
h_{0}=-\frac{\dot{\rho}_{m}(0)}{3\gamma_{m}\rho_{m}(0)},
\end{equation}
and
\begin{equation}\label{28}
h_{1}=\frac{1}{36h_{0}F_{RR}(0)}[{}\hspace{.1cm}3h_{0}^{2}F_{R}+\rho_{m}+\frac{1}{2}U(\phi)\dot{\phi}^{2}
-\frac{1}{2}F-3h_{0}F_{R\phi}\dot{\phi}{}\hspace{.1cm}]_{t=0}.
\end{equation}
So the parameter $h_{1}$ is, in general, different from zero,
which depends on initial values $\rho_{m}(0)$,
$\dot{\rho}_{m}(0)$, $\phi(0)$, and $\dot{\phi}(0)$, can be
positive or negative. This proves that, in general, the
generalized ST theories can explain the phantom-divide-line
crossing. The remaining two equations (\ref{23}) and (\ref{24})
determine the values of $\ddot{\phi}(0)$ and $\dddot{\phi}(0)$ in
terms of the four mentioned initial values. The higher order terms
of the expansions specify the other coefficients $h_{3}, h_{4},
\ldots$ For example eq.(\ref{21}) gives the parameter $h_{3}$.

It is clear that when the Lagrangian $F(\phi,R)$ is linear in $R$,
as is the case for scalar-tensor theories with action (\ref{2}),
then eq.(\ref{28}) is not the solution (note that in this case,
the last term of eq.(\ref{25}) vanishes). For these cases, it can
be shown that eq.(\ref{21}) specifics $h_{1}$ as follows:
\begin{equation}\begin{split}\label{29}
h_{1}=\frac{1}{4F_{R}(0)}[&3\rho_{m}\gamma_{m}^{2}h_{0}+6h_{0}U(\phi)\dot{\phi}^{2}-
F_{\phi}\dot{\phi}+F_{R\phi}(h_{0}\ddot{\phi}-\dddot{\phi})\\
&+F_{R\phi\phi}(h_{0}\dot{\phi}^{2}-3\dot{\phi}\ddot{\phi})-F_{R\phi\phi\phi}\dot{\phi}^{3}]_{t=0}.
\end{split}\end{equation}

For the next choice $\alpha=3$, it is clear from eq.(\ref{25})
that it does not lead to an expression for $h_{1}$, unlike the
case $\alpha=2$ in which eq.(\ref{25}) leads to (\ref{28}), but
instead, this equation leads to an extra relation between the
initial values $\rho_{m}(0)$, $\dot{\rho}_{m}(0)$, $\phi(0)$ and
$\dot{\phi}(0)$. For $\alpha=3$, eq.(\ref{26}) determines $h_{1}$.
So the $\alpha=3$ solution is also possible whenever there exists
a special relation between the four initial values. The situation
is worse for $\alpha\geq4$, since in these cases eq.(\ref{26})
also leads to another relation between the initial values. So
except for these fine-tuned initial values, the only acceptable
solution is $\alpha=2$, which always permits the $\omega=-1$
transition.

\section{The deceleration to acceleration transition}

To study the transition from $\ddot{a}<0$ to $\ddot{a}>0$, we
first note that
\begin{equation}\label{30}
\frac{\ddot{a}}{a}=\dot{H}+H^{2}.
\end{equation}
So at time $t'_{0}\equiv 0$ where $\ddot{a}=0$, one has
$\dot{H}=-H^{2}$. Expanding $H(t)$ around $t'_{0}=0$
\begin{equation}\label{31}
H(t)=H_{0}+H_{1}t+H_{2}t^{2}+\cdots
\end{equation}
then $\dot{H}=-H^{2}\biggm|_{t=0}$ results in
\begin{equation}\label{32}
H_{1}=-H_{0}^{2}.
\end{equation}
Similar to the previous section, one can seek the solutions for
eqs.(\ref{9}), (\ref{10}) and (\ref{14}), when $H(t)$ is given by
(\ref{31}), with condition (\ref{32}). In this case, the resulting
relations for $H_{0}$ and $H_{2}$ are
\begin{equation}\label{33}
H_{0}=-\frac{\dot{\rho}_{m}(0)}{3\gamma_{m}\rho_{m}(0)}
\end{equation}
and
\begin{equation}\label{34}
H_{2}=\frac{1}{36H_{0}F_{RR}(0)}\left[
{}\hspace{.1cm}\rho_{m}+\frac{1}{2}U(\phi)\dot{\phi}^{2}
+72H_{0}^{4}F_{RR}-3H_{0}F_{R\phi}\dot{\phi}-\frac{1}{2}F{}\hspace{.1cm}
\right]_{t=0},
\end{equation}
respectively. The other coefficients of expansion (\ref{31}) can
be also found consistently, which proves the existence of
$\ddot{a}=0$ crossing in generalized ST theories. For the cases
where $F(\phi,R)$ is linear in $R$, $H_{2}$ in eq.(\ref{34}) is
not more the solution and another expression, like eq.(\ref{29})
for $\omega=-1$ crossing, can be found.

\section{The coupled quintessence model}

As an explicit example, we consider the coupled quintessence model
introduced by following action:~\cite{faraoni}
\begin{equation}\label{35}
S=\frac{1}{2}\int {\rm d}^{4}x\sqrt{-g}{}\hspace{1ex}\left[
(1-\xi\phi^{2})R-(\nabla\phi)^{2} \right]+S_m,
\end{equation}
in which we do not consider the potential $V(\phi)$. At $\xi=0$,
eq.(\ref{35}) becomes the ordinary quintessence model, and for
$\xi\neq0$, the scalar field $\phi$ coupled nonminimally to the
gravity. The reason for considering the parameter $\xi$, first
introduced in~\cite{210faraoni}, has several answers which have
been discussed in~\cite{faraoni}. Perhaps the main reason to
include a $\xi\neq0$ term is that it is introduced by the first
loop corrections and is required by normalizability of the
theory~\cite{birrell}. If one considers a classical theory with
$\xi=0$, renormalization shifts it to one with $\xi\neq0$. We will
return to the quantum corrections in section 7.

In this section, we study the conditions under which the
$\omega=-1$ transition occurs in $V(\phi)=0$ coupled quintessence
model. For $\alpha=2$, our zero-order relations are
eqs.(\ref{22}), (\ref{23}) and (\ref{25}). Since here we will not
consider the matter field, for simplicity, the eq.(\ref{22}),
which is the Friedmann equation (\ref{10}) at zero order, is not
suitable. Instead we consider eq.(\ref{12}) and our three
independent Friedmann equations become eqs.(\ref{9}), (\ref{12})
and (\ref{14}), where for action (\ref{35}), result in:
\begin{equation}\label{36}
6H^{2}(1-\xi\phi^{2})-\dot{\phi}^{2}-12H\xi\phi\dot{\phi}-2\rho=0,
\end{equation}
\begin{equation}\label{37}
-2\dot{H}(1-\xi\phi^{2})-2H\xi\phi\dot{\phi}+2\xi\phi\ddot{\phi}
+(2\xi-1)\dot{\phi}^{2}=\rho+p,
\end{equation}
and
\begin{equation}\label{38}
\ddot{\phi}+3H\dot{\phi}+\xi\phi(6\dot{H}+12H^{2})=0,
\end{equation}
respectively. At zero order, these equations lead to:
\begin{equation}\label{39}
\dot{\phi}_{0}^{2}+12\xi
h_{0}\phi_{0}\dot{\phi}_{0}-6h_{0}^{2}(1-\xi\phi_{0}^{2})=0,
\end{equation}
\begin{equation}\label{40}
-2\xi
h_{0}\phi_{0}\dot{\phi}_{0}+2\xi\phi_{0}\ddot{\phi}_{0}+(2\xi-1)\dot{\phi}_{0}^{2}=0,
\end{equation}
\begin{equation}\label{41}
\ddot{\phi}_{0}+3h_{0}\dot{\phi}_{0}+12\xi h_{0}^{2}\phi_{0}=0,
\end{equation}
respectively. Here we assume that $S_{m}=0$. The first equation
specifies $\dot{\phi}_{0}$ in terms of $\phi_{0}$, and by
eliminating $\ddot{\phi}_{0}$ between eqs.(\ref{40}) and
(\ref{41}), one more expression for $\dot{\phi}_{0}$, in terms of
$\phi_{0}$, is obtained. These relations are
\begin{equation}\label{42}
\dot{\phi}_{0}=-h_{0}\left[
{}\hspace{.1cm}6\xi\phi_{0}\pm\sqrt{36\xi^{2}\phi_{0}^{2}
+6(1-\xi\phi_{0}^{2})}{}\hspace{.1cm}\right],
\end{equation}
and
\begin{equation}\label{43}
\dot{\phi}_{0}=-\frac{h_{0}}{1-2\xi}\left[{}\hspace{.1cm}4\xi
\phi_{0}\pm\xi\phi_{0}\sqrt{8(6\xi-1)}{}\hspace{.1cm}\right],
\end{equation}
respectively. Note that eq.(\ref{43}), which comes from
eqs.(\ref{37}) and (\ref{38}), holds only at the $\omega=-1$
transition point. This is because by setting $\dot{H}=0$, which is
the signature of $\omega=-1$ transition, eqs.(\ref{37}) and
(\ref{38}) lead to (\ref{40}) and (\ref{41}) and then (\ref{43})
is found. This is not the case for eq.(\ref{42}) which holds at
any instant of time.

The reality constraint of the scalar field $\phi$, restricts $\xi$
in eq.(\ref{43}) by
\begin{equation}\label{44}
\xi\geq\frac{1}{6}.
\end{equation}
Setting equal the eqs.(\ref{42}) and (\ref{43}), results in
\begin{equation}\label{45}
2\xi\phi_{0}\sqrt{6\xi-1}\left[{}\hspace{.1cm}\sqrt{6\xi-1}\pm\sqrt{2}{}\hspace{.1cm}\right]
=\pm(1-2\xi)\sqrt{36\xi^{2}\phi_{0}^{2}+6(1-\xi\phi_{0}^{2})}{}\hspace{.1cm}.
\end{equation}
Note that for $\xi\geq1/2$, the left-hand-side of eq.(\ref{45}) is
positive, so in the right-hand-side, only the minus sign must be
used for $\xi\geq1/2$. Solving $\phi_{0}$ from eq.(\ref{45}), one
finds
\begin{equation}\label{46}
{\phi_{0}^{(\pm)}}^{2}=\frac{3(1-2\xi)^{2}}{2\xi^{2}}\left[{}\hspace{.1cm}(6\xi-1)
\left(6\xi+1\pm2\sqrt{12\xi-2} \right)
-9(1-2\xi)^{2}+\frac{3(1-2\xi)^{2}}{2\xi}{}\hspace{.1cm}\right]^{-1}.
\end{equation}
It can be easily seen that ${\phi_{0}^{(-)}}^{2}$ is always
negative for all $\xi\geq1/6$, so there exists no real solution
for $\phi_{0}^{(-)}$. For ${\phi_{0}^{(+)}}^{2}$, it can be seen
that ${\phi_{0}^{(+)}}^{2}<0$ for $1/6\leq\xi\leq3/16$ and
${\phi_{0}^{(+)}}^{2}>0$ for $\xi>3/16$. So the real value of the
scalar field $\phi$ at transition time $t=0$, is $\phi_{0}^{(+)}$
from eq.(\ref{46}), and it exists if
\begin{equation}\label{47}
\xi>\frac{3}{16}.
\end{equation}
In other words, the coupled quintessence model can cross the
$\omega=-1$ line only when $\xi$ satisfies eq.(\ref{47}). This
leads us to consider the eq.(\ref{42}) and the right-hand-side of
eq.(\ref{45}) with "+" and "-" sign, for $3/16<\xi<1/2$ and
$\xi\geq1/2$, respectively\footnote{ Since only the plus sign of
eq.(\ref{46}), or eq.(\ref{43}) and the LHS of eq.(\ref{45}),
leads to acceptable solution for $\phi_0$, we must consider the
eq.(\ref{43}) and the LHS of eq.(\ref{45}) with $"+"$ sign. Under
this condition, the LHS of eq.(\ref{45}) becomes positive for
$3/16<\xi<1/2$, therefore the $"+"$ sign must be chosen in the RHS
of this equation and therefore in eq.(\ref{42}). For $\xi\geq1/2$,
the accepted sign in the RHS of eq.(\ref{45}) and in eq.(\ref{42})
is $"-"$, as pointed out after eq.(\ref{45}).}.

To determine the crossing behavior of this models, we must
calculate the parameter $h_{1}$ from eq.(\ref{29}), or instead,
from perturbative calculation of eq.(\ref{12}), or (\ref{37}). To
do so, it is better to rewrite the differential
eqs.(\ref{36})-(\ref{38}) in terms of red-shift parameter $z$,
defined through
\begin{equation}\label{48}
1+z=\frac{a_{*}}{a},
\end{equation}
instead of the time variable $t$. In eq.(\ref{48}), $a_{*}$ is the
scale factor of the universe in the present time. Using
\begin{equation}\begin{split}\label{49}
&\frac{\rm d}{{\rm d}t}=-H(1+z)\frac{\rm d}{{\rm
d}z}{}\hspace{.2cm},\\
&\frac{\rm d^{2}}{{\rm d}t^{2}}=H^{2}(1+z)^{2}\frac{\rm
d^{2}}{{\rm d}z^{2}}+H(1+z)^{2}\frac{{\rm d}H}{{\rm d}z}\frac{\rm
d}{{\rm d}z}+H^{2}(1+z)\frac{\rm d}{{\rm d}z},
\end{split}\end{equation}
the first Friedmann equation (\ref{36}) results in $\eta(z)={\rm
d}\phi/{\rm d}z$ in terms of $\phi(z)$, and eqs.(\ref{37}) and
(\ref{38}) specify ${\rm d}H/{\rm d}z$ and ${\rm d}\eta/{\rm d}z$.
The resulting equations are:
\begin{equation}\label{50}
\eta(z)=\frac{{\rm d}\phi(z)}{{\rm d}z}{}\hspace{.2cm},
\end{equation}
\begin{equation}\label{51}
(1+z)\eta(z)=6\xi\phi(z)\pm\sqrt{36\xi^{2}\phi^{2}(z)+6(1-\xi\phi^{2}(z))}{}\hspace{.2cm},
\end{equation}

\begin{equation}\label{52}
\frac{{\rm d}H(z)}{{\rm
d}z}=\frac{24\xi^{2}\phi^{2}(z)+(1-2\xi)(1+z)^{2}\eta^{2}(z)
-8\xi(1+z)\eta(z)\phi(z)}{2(1+z)[{}\hspace{.1cm}1+(6\xi^{2}-\xi)\phi^{2}(z){}\hspace{.1cm}]}H(z){}\hspace{.2cm},
\end{equation}

\begin{equation}\begin{split}\label{53}
&\frac{{\rm d}\eta(z)}{{\rm d}z}= \frac{1}
{2(1+z)[{}\hspace{.1cm}1+(6\xi^{2}-\xi)\phi^{2}(z){}\hspace{.1cm}]}
\times \left\{ 24\xi^{2}\phi^{3}(z)+(2\xi-1)(1+z)^{3}\eta^{3}(z)
\right.
\\
&\left.
+4(1+z)[{}\hspace{.1cm}1-(\xi+12\xi^{2})\phi^{2}(z){}\hspace{.1cm}]\eta(z)
+2(1+z)^{2}(7\xi-6\xi^{2})\eta^{2}(z)\phi(z)-24\xi\phi(z)\right\}.
\end{split}\end{equation}
As explained after eq.(\ref{47}), for $3/16<\xi<1/2$,
eq.(\ref{51}) must be used with $"+"$ sign and for $\xi\geq 1/2$,
the $"-"$ sign is accepted.

Expanding $H(z)$ around the transition point $z_{0}$, one finds
\begin{equation}\label{54}
H(z)=h_{0}+h'_{1}(z-z_{0})^{2}+h'_{2}(z-z_{0})^{3}+\cdots{}\hspace{.2cm}.
\end{equation}
in which we choose $\alpha=2$. Note that
$\omega=-1-(2/3)\dot{H}/H^{2}=-1+\frac{2(1+z)}{3H}{\rm d}H/{\rm
d}z$, so at transition point $z_{0}$, one has ${\rm d}H/{\rm
d}z|_{z=z_0}=0$. It is clear from eq.(\ref{52}) that ${\rm
d}H/{\rm d}z=A(z)H(z)$, so $A(z_{0})=0$ from which ${{{\rm
d}^{2}}H/{\rm d}z^{2}}|_{z=z_0}=H(z){{\rm d}A/{\rm d}z}|_{z=z_0}$.
In this way, eqs.(\ref{50})-(\ref{53}) finally result in
$h'_{1}=(1/2){{{\rm d}^{2}}H/{\rm d}z^{2}}|_{z=z_0}$ as follows
\begin{equation}\begin{split}\label{55}
&h'_{1}=\frac{h_{0}}{8(1+z_{0})^{2}[{}\hspace{.1cm}1+(6\xi^{2}-\xi)\phi^{2}_{0}{}\hspace{.1cm}]^2}\times
\{96(1+z_{0})[{}\hspace{.1cm}(6\xi^{4}+\xi^{3}+\xi^{2})\phi^{2}_{0}+2\xi^{2}-\xi{}\hspace{.1cm}]\phi_{0}\eta_{0}\\
&+4(1+z_{0})^{2}[{}\hspace{.1cm}(36\xi^{3}-36\xi^{2}-3\xi)\phi^{2}_{0}-10\xi
+3{}\hspace{.1cm}]\eta^{2}_{0}+12(1+z_{0})^{3}(4\xi^{3}-6\xi^{2}+3\xi)\phi_{0}\eta^{3}_{0}\\
&-2(1+z_{0})^{4}(4\xi^{2}-4\xi+1)\eta^{4}_{0}+192\xi^{2}\phi^{2}_{0}-192\xi^{3}\eta^{4}_{0}\}.
\end{split}\end{equation}
In this equation, $\phi_{0}=\phi(z_{0})$ and
$\eta_{0}=\eta(z_{0})$.

At the first step, we solve numerically the differential equations
(\ref{50}), (\ref{52}) and (\ref{53}) for the specific example
$\xi=1$. By choosing the initial values $\phi(z=0)=1/2$ and
$H(z=0)=72$ and calculating $\eta(z=0)$ from eq.(\ref{51}) with
$"-"$ sign, because $\xi>1/2$, one finds $H(z),\phi(z),\eta(z)$
and $\omega(z)=-1+\frac{2(1+z)}{3H}{\rm d}H/{\rm d}z$ in
Figs.(1)-(4).
\begin{figure}[hbt]\label{Figure 1}
\centering
\hspace{2cm}\includegraphics*[height=5cm,width=7cm]{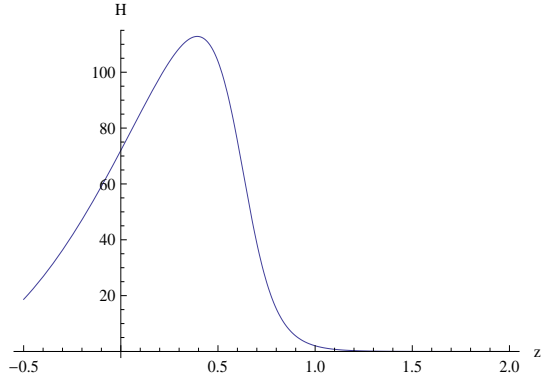}\hspace{2cm}
\caption{ The plot of $H(z)$ of coupled quintessence model with
$\xi=1$.}
\end{figure}
\begin{figure}[hbt]\label{Figure_2}
\centering
\hspace{2cm}\includegraphics*[height=5cm,width=7cm]{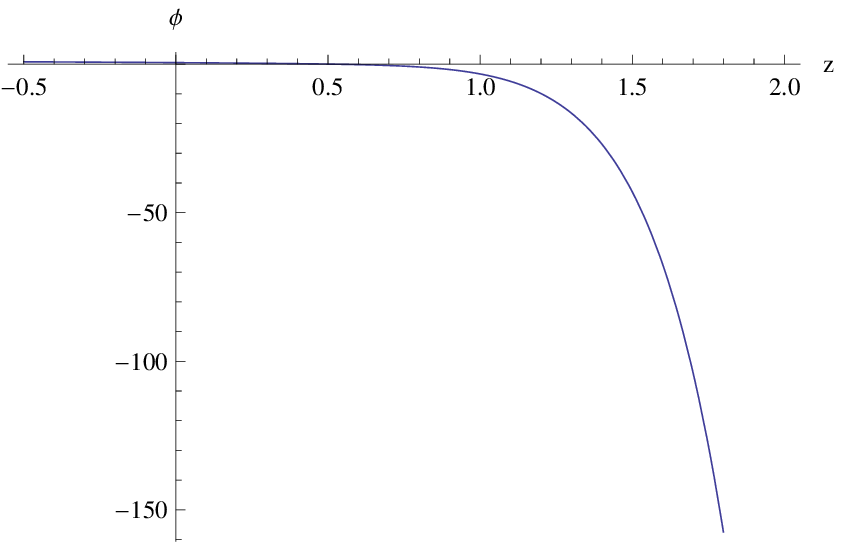}\hspace{2cm}
\caption{ The plot of $\phi(z)$ of coupled quintessence model with
$\xi=1$.}
\end{figure}
\begin{figure}[hbt]\label{Figure 3}
\centering
\hspace{2cm}\includegraphics*[height=5cm,width=7cm]{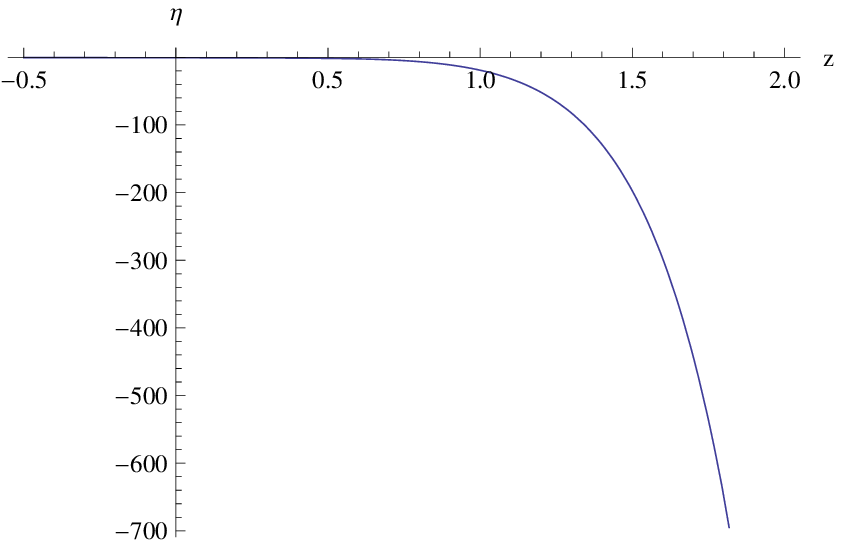}\hspace{2cm}
\caption{ The plot of $\eta(z)={\rm d}\phi(z)/{\rm d}z$ of coupled
quintessence model with $\xi=1$.}
\end{figure}
\begin{figure}[hbt]\label{Figure_4}
\centering
\hspace{2cm}\includegraphics*[height=5cm,width=7cm]{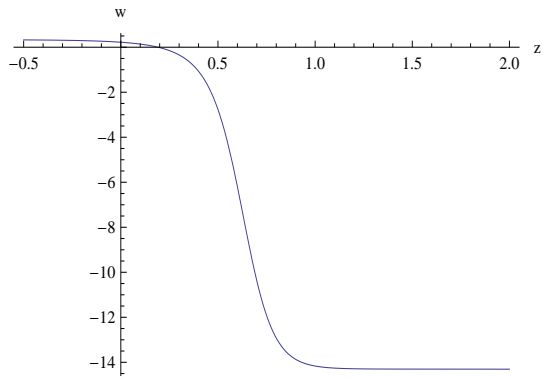}\hspace{2cm}
\caption{ The plot of $\omega(z)$ of coupled quintessence model
with $\xi=1$. It crosses $\omega=-1$ at $z_{0}=0.393$.}
\end{figure}

As it is clear from Fig.(4), $\omega(z)$ crosses the $\omega=-1$
line, from $\omega>-1$ region to $\omega<-1$ region, at
$z_{0}=0.393$. At this point, $H(z_{0})=112.83$,
$\phi(z_{0})=0.16$, and $\eta(z_{0})=-1.18$.

Our perturbative results verify this crossing. For $\xi=1,
\phi^{(+)}_{0}$ from eq.(\ref{46}) becomes $\phi^{(+)}_{0}=0.16$,
which is in coincidence with numerical calculations. Choosing
$z_{0}=0.393$ and $h_{0}=112.83$, eq.(\ref{51}) results in
$\eta(z_{0})=-1.18$, and $h'_{1}$ from eq.(\ref{55}) becomes
$h'_{1}=-618.9$. Therefore, up to lowest order, we have:
\begin{equation}\begin{split}\label{56}
&\omega(z)=-1+\frac{2(1+z)}{3H(z)}\frac{{\rm d}H}{{\rm
d}z}=-1+\frac{4h'_{1}}{3h_{0}}(1+z)(z-z_{0})+\cdots\\
&=-1-7.31(1+z)(z-0.393)+\cdots
\end{split}\end{equation}
Fig.(5) shows the relation (\ref{56}), which has the similar
behavior, near the transition point $z=z_{0}$, as the Fig.(4).
Note that in both approaches, the number of needed initial values
are the same. In numerical calculation, we use two initial values
$\phi(z=0)$ and $H(z=0)$, and in perturbative calculation, two
parameters $z_{0}$ and $h_{0}=H(z=z_{0})$ are chosen.
\begin{figure}[hbt]\label{Figure_5}
\centering
\hspace{2cm}\includegraphics*[height=5cm,width=7cm]{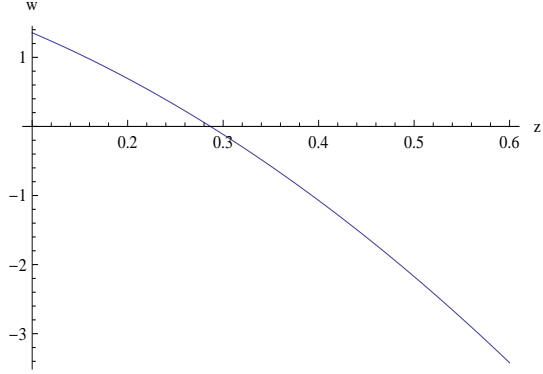}\hspace{2cm}
\caption{ The perturbative variation of $\omega(z)$ of coupled
quintessence model with $\xi=1$.}
\end{figure}

\section{The Brans-Dicke model}

As a second example, we consider the Jordan-Brans-Dicke model with
$V(\phi)=0$ and $S_{m}=0$,
\begin{equation}\label{57}
S=\frac{1}{2}\int {\rm d}^{4}x\sqrt{-g}{}\hspace{1ex}\left[ \phi
R-\frac{{\rm{w}}}{\phi}(\nabla\phi)^2 \right].
\end{equation}
Comparing (\ref{57}) with (\ref{3}), gives $F(\phi)=R\phi$ and
$U(\phi)={\rm{w}}/\phi$. In this way, eqs.(\ref{23})-(\ref{26}),
with $\alpha=2$, and (\ref{29}) become:
\begin{equation}\label{58}
12h^{2}_{0}+{\rm{w}}A^{2}-2{\rm{w}}(3h_{0}A+B)=0,
\end{equation}
\begin{equation}\label{59}
-C-A^{3}+2AB+3h_{0}A^{2}-3h_{0}B+\frac{6}{{\rm{w}}}h_{1}=0,
\end{equation}
\begin{equation}\label{60}
-{\rm{w}}A^{2}+6h_{0}A+6h^{2}_{0}-2\frac{\rho_0}{\phi_0}=0,
\end{equation}
\begin{equation}\label{61}
B+{\rm{w}}A^{2}-h_{0}A+\frac{\rho_0+p_0}{\phi_{0}}=0,
\end{equation}
and
\begin{equation}\label{62}
h_{1}=\frac{1}{4}\left[\hspace{.1cm}-C+h_{0}B+6h_{0}{\rm{w}}A^{2}
-12h^{2}_{0}A+\frac{3\rho_0\gamma^{2}h_{0}}{\phi_{0}}\hspace{.1cm}\right],
\end{equation}
respectively. In above equations, the parameters $A, B$, and $C$
are
\begin{equation}\label{63}
A=\left(\frac{\dot{\phi}}{\phi}\right)_{0}\hspace{.5cm},\hspace{.5cm}B=\left(\frac{\ddot{\phi}}{\phi}\right)_{0}
\hspace{.5cm},\hspace{.5cm}C=\left(\frac{\dddot{\phi}}{\phi}\right)_{0}\hspace{.1cm},
\end{equation}
and $R(0)=12h^{2}_{0}$ has been used. Setting $\rho_0=p_0=0$ in
eqs.(\ref{58})-(\ref{62}), to ensure $S_{m}=0$, eq.(\ref{60}) then
gives $A$ as follows:
\begin{equation}\label{64}
A=\frac{3h_{0}}{{\rm{w}}}\left[\hspace{.1cm}1\pm\sqrt{1+\frac{2{\rm{w}}}{3}}\hspace{.1cm}\right].
\end{equation}
Reality of $A$ demands
\begin{equation}\label{64'}
{\rm{w}}>-\frac{3}{2},
\end{equation}
where for the reasons which will be discussed later, we do not
consider the ${\rm{w}}=-3/2$ case.

The parameter $B$ can be found from either eqs.(\ref{58}) or
(\ref{61}). Using eq.(\ref{60}), they yield
\begin{equation}\label{65}
B=-(5h_{0}A+6h^{2}_{0})=\frac{9}{{\rm{w}}}h^{2}_{0}+3(\frac{1}{{\rm{w}}}-1)h_{0}A.
\end{equation}
Putting (\ref{64}) into eq.(\ref{65}), the parameter 'w' is fixed
as
\begin{equation}\label{66}
{\rm{w}}=-\frac{4}{3}.
\end{equation}
In other words, the Brans-Dicke theory has $\omega=-1$ only when
${\rm{w}}=-4/3$. Putting (\ref{66}) back into eq.(\ref{64}), with
plus sign\footnote{ Note that for the minus sign of eq.(\ref{64}),
the solution of eq.(\ref{65}) is ${\rm{w}}=-3/2$, and for the plus
sign it is ${\rm{w}}=-3/2$ and ${\rm{w}}=-4/3$. Because of the
physical condition (\ref{64'}), ${\rm{w}}=-4/3$ is the only
acceptable solution and therefore we only consider the plus sign
of eq.(\ref{64}).}, and eq.(\ref{65}), results in
\begin{equation}\label{67}
A=-3h_{0}\hspace{.5cm},\hspace{.5cm}B=9h^{2}_{0}\hspace{.1cm}.
\end{equation}
The parameter $h_{1}$ can be found by either eqs.(\ref{59}) and
(\ref{62}), where using (\ref{66}) and (\ref{67}), it gives the
result
\begin{equation}\label{68}
h_{1}=-\frac{2}{9}(C+27h^{3}_{0})=-\frac{1}{4}(C+27h^{3}_{0}).
\end{equation}
The above equality gives $C$:
\begin{equation}\label{69}
C=-27h^{3}_{0},
\end{equation}
from which
\begin{equation}\label{70}
h_{1}=0.
\end{equation}
So it seems that there is no $\omega=-1$ crossing phenomenon in
Brans-Dicke model. In fact, calculating the other coefficients of
expansion (\ref{16}) results in $h_{i}(i\geq1)=0$.

To justify our result, let us exactly calculate the equation of
state parameter $\omega$ of the Brans-Dicke model. We first write
eqs.(\ref{9}), (\ref{12}) and (\ref{14}) for this model, which
results in
\begin{equation}\label{71}
3H^{2}\phi=\frac{1}{2}\frac{{\rm{w}}}{\phi}\dot{\phi}^{2}-3H\dot{\phi},
\end{equation}
\begin{equation}\label{72}
-2\dot{H}\phi+H\dot{\phi}-\ddot{\phi}-\frac{{\rm{w}}}{\phi}\dot{\phi}^{2}=0,
\end{equation}
and
\begin{equation}\label{73}
6(\dot{H}+2H^{2})-2\frac{{\rm{w}}}{\phi}(\ddot{\phi}+3H\dot{\phi})+\frac{{\rm{w}}}{\phi^{2}}\dot{\phi}^{2}=0,
\end{equation}
respectively. Using the red-shift parameter $z$, the above
equations can be used to obtain the following relations in
$z$-space:
\begin{equation}\label{74}
6={\rm{w}}(1+z)^{2}\left(\frac{\eta(z)}{\phi(z)}\right)^{2}+6(1+z)\frac{\eta(z)}{\phi(z)},
\end{equation}
\begin{equation}\label{75}
\frac{{\rm d}H(z)}{{\rm
d}z}=\frac{H(z)}{(1+z)(6+4{\rm{w}})}\left[\hspace{.1cm}{\rm{w}}(1+2\rm{w})(1+z)^{2}\left(
\frac{\eta(z)}{\phi(z)}\right)^{2}
+8{\rm{w}}(1+z)\frac{\eta(z)}{\phi(z)}+12\hspace{.1cm}\right],
\end{equation}
and
\begin{equation}\begin{split}\label{76}
&\frac{{\rm d}\eta(z)}{{\rm d}z}=\frac{\phi
(z)}{(1+z)(6+4\rm{w})}\left[\hspace{.1cm}-{\rm{w}}(1+2{\rm{w}})(1+z)^{2}
\frac{\eta^{3}(z)}{\phi^{3}(z)}\right.\\
&\left.
-12{\rm{w}}(1+z)\frac{\eta^{2}(z)}{\phi^2(z)}+8({\rm{w}}-3)
\frac{\eta(z)}{\phi(z)}+\frac{24}{1+z}\hspace{.1cm}\right],
\end{split}\end{equation}
in which $\eta(z)={\rm d}\phi(z)/{\rm d}z$. Because of the
denominator of eqs.(\ref{75}) and (\ref{76}), we take
${\rm{w}}\neq-3/2$ for these equations. We will consider the
${\rm{w}}=-3/2$ case separately. Solving eq.(\ref{74}) for
$\eta(z)/\phi(z)$, results in
\begin{equation}\label{77}
\frac{\eta(z)}{\phi(z)}=\frac{1}{{\rm{w}}(1+z)}[\hspace{.1cm}-3\pm\sqrt{9+6{\rm{w}}}\hspace{.1cm}],
\end{equation}
in which we assume ${\rm{w}}\neq0$ and also the condition
(\ref{64'}). Putting eq.(\ref{77}) into eq.(\ref{75}), gives
\begin{equation}\label{78}
\frac{{\rm d}H(z)}{{\rm d}z}=\frac{g_{\mp}({\rm{w}})}{1+z}H(z),
\end{equation}
where
\begin{equation}\label{79}
g_{\mp}({\rm{w}})=\frac{1}{{\rm{w}}}[\hspace{.1cm}3({\rm{w}}+1)\mp\sqrt{3(2{\rm{w}}+3)}\hspace{.1cm}].
\end{equation}
In this way the exact solution of the Brans-Dicke model is found
as follows:
\begin{equation}\label{80}
H(z)=H(z_{0})(1+z)^{g_{\mp}({\rm{w}})}\hspace{.1cm}.
\end{equation}
The equation of state parameter $\omega$ can be found as
\begin{equation}\label{81}
\omega(z)=-1+\frac{2}{3}\frac{(1+z)}{H(z)}\frac{{\rm d}H}{{\rm
d}z}=-1+\frac{2}{3}g_{\mp}({\rm{w}}),
\end{equation}
which is constant in time. So it is natural that it does not cross
the phantom-divide-line, and $\omega(z)=-1$ occurs when
$g_{\mp}({\rm{w}})=0$. $g_{-}({\rm{w}})=0$ results in ${\rm{w}}=0$
which is not acceptable, and $g_{+}(\rm{w})=0$ results in
${\rm{w}}=-4/3$ which is in accordance with the perturbative
result obtained in eq.(\ref{66}).

It is interesting to note that our exact solution (\ref{80})
coincides with O'Hanlon and Tupper solution:~\cite{ohanlon}
\begin{equation}\label{82}
a(t)=a_{0}\left(\frac{t}{t_{0}}\right)^{q_{\pm}},
\end{equation}
\begin{equation}\label{83}
\phi(t)=\phi_{0}\left(\frac{t}{t_{0}}\right)^{s_{\pm}},
\end{equation}
with
\begin{equation}\begin{split}\label{84}
&q_{\pm}=\frac{{\rm{w}}}{3({\rm{w}}+1)\mp\sqrt{3(2{\rm{w}}+3)}}\hspace{.2cm},\\
&s_{\pm}=\frac{1\mp\sqrt{3(2{\rm{w}}+3)}}{3{\rm{w}}+4}\hspace{.2cm}.
\end{split}\end{equation}
In $t$-space, (\ref{82}) results in
\begin{equation}\label{85}
H(t)=\frac{q_{\pm}}{t}\Rightarrow\hspace{.2cm}\omega=-1-\frac{2}{3}\frac{\dot{H}}{H^{2}}
=-1+\frac{2}{3q_{\pm}},
\end{equation}
which by noting that $g_{\mp}=1/q_{\pm}$ (see eqs.(\ref{79}) and
(\ref{84})), the $\omega$ in (\ref{85}) becomes the same as one in
eq.(\ref{81}).

The only remaining case is ${\rm{w}}=-3/2$. To see the behavior of
the Brans-Dicke model at ${\rm{w}}=-3/2$, we consider the
Friedmann equations (\ref{71})-(\ref{73}). From the first equation
(\ref{71}), one finds
\begin{equation}\label{86}
\frac{1}{H}\frac{\dot{\phi}}{\phi}=\frac{3\pm\sqrt{9+6{\rm{w}}}}{\rm{w}}\biggm|_{{\rm{w}}=-\frac{3}{2}}=-2,
\end{equation}
from which
\begin{equation}\begin{split}\label{87}
&\dot{\phi}=-2H\phi\hspace{.2cm},\\
&\ddot{\phi}=-2\dot{H}\phi+4H^{2}\phi\hspace{.2cm}.
\end{split}\end{equation}
Putting these equations back into the second and third Friedmann
equations (\ref{72}) and (\ref{73}), which are not independent at
${\rm{w}}=-3/2$, it can be easily seen that these equations
trivially hold for arbitrary $H(t)$ function. Therefore we exclude
${\rm{w}}=-3/2$ case in which all the $H(t)$ functions are
solutions of the Friedmann equations.

\section{The quantum corrections}

As it was discussed earlier, the classical Lagrangian
\begin{equation}\label{88}
\mathcal{L}=\frac{1}{2}(-\xi R\phi^{2}-(\nabla\phi)^{2})
\end{equation}
can induce interesting quantum effects. The renormalizability
requirements enforces $\xi$ in the above action to be $\xi=1/6$.
In $\xi=1/6$, the model has several other interesting properties,
such as the conformal invariance, holding the Einstein equivalence
principle, etc.~\cite{faraoni,birrell}.

Calculating the effective action of this model at one-loop level,
results in some extra terms in the trace of the energy-momentum
tensor, which is trace-less classically. For this reason, this
effect is called the trace/conformal anomaly. These extra terms
are~\cite{starobinsky,birrell}.
\begin{equation}\label{89}
T=b(F+\frac{2}{3}\Box R)+b'G+b''\Box R\hspace{.2cm},
\end{equation}
where $T$ denotes the trace of the energy-momentum tensor, $F$ is
the square of the 4d Weyl tensor and $G$ is the Gauss-Bonnet
invariant
\begin{equation}\begin{split}\label{90}
&F=\frac{1}{3}R^{2}-2R_{\mu\nu}R^{\mu\nu}
+R_{\mu\nu\alpha\beta}R^{\mu\nu\alpha\beta}\hspace{.2cm},\\
&G=R^{2}-4R_{\mu\nu}R^{\mu\nu}+R_{\mu\nu\alpha\beta}R^{\mu\nu\alpha\beta}\hspace{.2cm}.
\end{split}\end{equation}
For $N$ scalars, $N_{1/2}$ spinors, $N_{1}$ vector fields,
$N_{2}$(=0 or 1) gravitons and $N_{HD}$ higher derivative
conformal scalars, $b$, $b'$ and $b''$ are given by
\begin{equation}\begin{split}\label{91}
&b=\frac{N+6N_{1/2}+12N_1+611N_2-8N_{HD}}{120(4\pi)^2}\hspace{.2cm},\\
&b'=-\frac{N+11N_{1/2}+62N_1+1411N_2-28N_{HD}}{360(4\pi)^2}\hspace{.2cm},\hspace{.2cm}
b''=0.
\end{split}\end{equation}
Using eq.(\ref{89}) for the case of FRW metric, it can be shown
that the contribution of conformal anomaly to energy density and
pressure are~\cite{nojiri4}
\begin{equation}\begin{split}\label{92}
&\rho_{A}=-\frac{1}{a^{4}}\{b'(6a^{4}H^{4}+12a^{2}H^{2})\\
&+(\frac{2}{3}b+b'')[\hspace{.1cm}a^{4}(-6H\ddot{H}-18H^{2}\dot{H}+3\dot{H}^{2})
+6a^{2}H^{2}\hspace{.1cm}]\\
&-2b+6b'-3b''\}\hspace{.2cm},
\end{split}\end{equation}
and
\begin{equation}\begin{split}\label{93}
&p_{A}=b'[\hspace{.1cm}6H^{4}+8H^{2}\dot{H}
+\frac{1}{a^{2}}(4H^{2}+8\dot{H})\hspace{.1cm}]\\
&+(\frac{2}{3}b+b'')[\hspace{.1cm}-2\dddot{H}-12H\ddot{H}
-18H^{2}\dot{H}-9\dot{H}^{2}+\frac{1}{a^{2}}(2H^{2}+4\dot{H})\hspace{.1cm}]\\
&-\frac{-2b+6b'-3b''}{3a^{4}}.
\end{split}\end{equation}
The subscript $"A"$ stands for "anomaly". To consider the quantum
effects in gravitational phenomena, one can add the above
$\rho_{A}$ and $p_{A}$ to the Friedmann equations. In this way the
contribution of quantum effects on $\omega=-1$ crossing can be
studied.

\subsection{The coupled-quintessence model}

We first study how much the adding of the quantum terms $\rho_{A}$
and $p_{A}$ can change the $\omega=-1$ crossing of the coupled
quintessence model discussed in section 5. As this model is a
theory with one scalar field, one has $N=1$ and
$N_{1/2}=N_{1}=N_{2}=N_{HD}=0$. Eq.(\ref{91}) then results in
\begin{equation}\label{94}
b=-3b'=\frac{1}{120(4\pi)^{2}}.
\end{equation}

The Friedmann equations of coupled quintessence model are
eqs.(\ref{36})-(\ref{38}), in which we put $\rho_{A}$ and $p_{A}$
as the source terms. At zero-order, eqs.(\ref{92}) and (\ref{93}),
using (\ref{94}), are
\begin{equation}\label{95}
\rho_{A}(0)=-b'(8h^{4}_{0}+24h_{0}h_{1}+\frac{12}{a^{4}_{0}}),
\end{equation}
and
\begin{equation}\label{96}
p_{A}(0)=b'(6h^{4}_{0}+48h_{0}h_{1}-\frac{4}{a^{4}_{0}}+24h_{2}),
\end{equation}
respectively. Therefore, instead of eqs.(\ref{39}) and (\ref{40}),
now we have:
\begin{equation}\label{97}
\dot{\phi}^{2}_{0}+12\xi
h_{0}\phi_{0}\dot{\phi}_{0}-6h^{2}_{0}(1-\xi\phi^{2}_{0})
-2b'(8h^{4}_{0}+24h_{0}h_{1}+\frac{12}{a^{4}_{0}})=0\hspace{.2cm},
\end{equation}
and
\begin{equation}\label{98}
-2\xi h_{0}\phi_{0}\dot{\phi}_{0}+2\xi\phi_{0}\ddot{\phi}_{0}
+(2\xi-1)\dot{\phi}^{2}_{0}=b'(-2h^{4}_{0}+24h_{0}h_{1}
-\frac{16}{a^{4}_{0}}+24h_{2})\hspace{.2cm},
\end{equation}
respectively. Eq.(\ref{41}) is not changed. As it is clear from
the above equations, the situation is very different from
classical relations (\ref{39})-(\ref{41}). Here we do not have
the relations like eqs.(\ref{42}) and (\ref{43}) which describe
$\dot{\phi}_{0}$ in terms of $\phi_{0}$, and therefore the
constraint like eq.(\ref{47}) does not appear. In other words, the
$\omega=-1$ crossing can happen for any $\xi$ values. In fact,
eq.(\ref{97}) determines $h_{1}$ as follows:
\begin{equation}\label{99}
h_{1}=\frac{\dot{\phi}^{2}_{0}+12\xi h_{0}\phi_{0}\dot{\phi}_{0}
-6h^{2}_{0}(1-\xi\phi^{2}_{0})-8b'(2h^{4}_{0}
+3/a^4_0)}{48b'h_{0}}\hspace{.2cm},
\end{equation}
which is generally different from zero for any $\xi$ values,
including $\xi=1/6$. In this way we obtain an interesting result:
The coupled quintessence model classically crosses the $\omega=-1$
line for $\xi>3/16$, but because of the quantum effects, the
$\xi=1/6<3/16$ is also allowed. It is interesting to note that the
quantum correction terms (the last two terms in the numerator of
eq.(\ref{99})) are much smaller than the classical terms, since
\begin{equation}\label{100}
h^{4}_{0}\sim\frac{1}{a^{4}_{0}}\ll h^{2}_{0}.
\end{equation}

\subsection{The Brans-Dicke model}

In this case, the quantum induced energy density and pressure
(\ref{95}) and (\ref{96}) must be inserted into
eqs.(\ref{58})-(\ref{62}), in order to find the influence of
quantum phenomena on $\omega=-1$ transition of Brans-Dicke model.
Eq.(\ref{60}) then becomes
\begin{equation}\label{101}
-{\rm{w}}\frac{\dot{\phi}^{2}_{0}}{\phi_{0}}+6h_{0}\dot{\phi}_{0}
+6h^{2}_{0}\phi_{0}+2b'(8h^{4}_{0}+24h_{0}h_{1}+\frac{12}{a^{4}_{0}})=0.
\end{equation}
In contrast to classical case in which eq.(\ref{60}), with
$\rho_0=0$, specifies $(\dot{\phi}/\phi)_{0}$ in terms of "w" in
eq.(\ref{64}), from which ${\rm{w}}=-4/3$ becomes necessary for
achieving $\omega=-1$, see eq.(\ref{66}), and finally results in
$h_{1}=0$ in eq.(\ref{70}), here the eq.(\ref{60}), in its new
form (\ref{101}), results in a nonzero $h_{1}$:
\begin{equation}\label{102}
h_{1}=\frac{{\rm{w}}\dot{\phi}^{2}_{0}/\phi_{0}-6h_{0}\dot{\phi}_{0}
-6h^{2}_{0}\phi_{0}-8b'(2h^{4}_{0}+3/a^{4}_{0})}{48b'h_{0}}.
\end{equation}
This shows that the quantum effects can produce the non-zero
$h_{1}$ which indicates the existence of $\omega=-1$ crossing.

There is a last point which must be noted. Comparing the
Brans-Dicke Lagrangian (\ref{57}) with Lagrangian (\ref{88}),
which $\rho_{A}$ and $p_{A}$ in eqs.(\ref{95}) and (\ref{96}) are
derived from, may lead us to conclude that it is not reasonable to
use $\rho_{A}$ and $p_{A}$ as the quantum corrections of
(\ref{57}). This is because these two Lagrangians are different,
so the corrections come from (\ref{88}) can not be used for
(\ref{57}). The answer is that if we apply the following change of
field variable to eq.(\ref{57}):
\begin{equation}\label{103}
\phi\rightarrow\frac{1}{4{\rm{w}}}\phi^{2},
\end{equation}
the Brans-Dicke action then becomes
\begin{equation}\label{104}
S=\frac{1}{2}\int {\rm d}^{4}x\sqrt{-g}{}\hspace{1ex}\left(
\frac{1}{4{\rm{w}}}R\phi^{2}-(\nabla\phi)^2 \right)\hspace{.1cm}.
\end{equation}
By choosing $\xi=-1/4\rm{w}$, eq.(\ref{104}) becomes the same as
the Lagrangian (\ref{88}), so its quantum computations can be used
for Brans-Dicke action. Restricting ourselves to $\xi=1/6$,
results in ${\rm{w}}=-3/2$, which is not in the region of our
interest. So in the physical region ${\rm{w}}>-3/2$, the quantum
phenomena can not change the noncrossing behavior of equation of
state parameter from $\omega=-1$ line.

\subsection{Deceleration to acceleration transition}
To study the influences of quantum effects on $\ddot{a}=0$
transition, the quantum terms $\rho_A$ and $p_A$ must be added to
Friedmann equations (\ref{9}), (\ref{10}) and (\ref{14}), and
solved for the coefficients of the expansion $H(t)$ in
eq.(\ref{31}). Eq.(\ref{33}) does not change, but the coefficient
$H_2$ in eq.(\ref{34}), for $b=-3b'$ case, is replaced by
\begin{equation}\begin{split}\label{106}
H_{2}=\frac{1}{12H_{0}(2b'+3F_{RR}(0))}[
{}\hspace{.1cm}&\rho_{m}+\frac{1}{2}U(\phi)\dot{\phi}^{2}
+72H_{0}^{4}F_{RR}-3H_{0}F_{R\phi}\dot{\phi}\\
& -\frac{1}{2}F{}+
12b'(3H_0^4-\frac{1}{a_0^4})\hspace{.1cm}]_{t=0}.
\end{split}\end{equation}
In above equation, $a_0$ is the scale factor $a(t)$ at the
transition time. It is seen that in $b'\rightarrow 0$ limit,
eq.(\ref{106}) reduces to classical relation (\ref{34}), as it is
expected. For special cases coupled quintessence and Brans-Dicke
models, $H_2$ can be easily calculated from eq.(\ref{106}).

\section{Conclusion}

In this paper, we consider the generalized scalar-tensor models as
a class of modified gravity theories to describe two important
cosmological transitions, the $\omega=-1$ and $\ddot{a}=0$
transitions. For both cases, we solve the Friedmann equations of
generalized ST models by perturbative expansion of the Hubble
parameter around the transition points. The expansion parameters
are found consistently, which proves the existence of these
solutions.

Two specific examples are studied in details. The first one is the
coupled quintessence model, which the scalar field is coupled to
gravity nonminimally. It is shown that for all $\xi>3/16$ cases,
the model has a $\omega=-1$ transition, a fact which can be seen
by numerical solving of the Friedmann equations for several
$\xi$-values. The $\xi=1$ results are reported. The second example
is the Brans-Dicke model, which our perturbative method shows no
transition, a fact that is verified by solving exactly the
Friedmann equations of this model. It is seen that the equation of
state parameter of this model is constant, which naturally never
cross the $\omega=-1$ line. Our investigation shows that "w" must
satisfy ${\rm{w}}>-3/2$.

Finally we consider the quantum terms coming from the one-loop
calculation of a scalar field, coupled non-minimally to gravity.
It is shown that these quantum field theoretical terms can, in
general, change the $\omega=-1$ and $\ddot{a}=0$ crossing
behaviors of the models. For instance it is shown that the coupled
quintessence model with $\xi=1/6$, which classically can not cross
the $\omega=-1$ line, crosses this line because of quantum
effects. For Brans-Dicke model, this phenomenon has no effect on
transition of models with ${\rm{w}}>-3/2$.

{\bf Acknowledgement:} This work was partially supported by the
"center of excellence in structure of matter" of the Department of
Physics of the University of Tehran, and also a research grant
from the University of Tehran\\ \\

\end{document}